\documentclass[conference,onecolumn]{IEEEtran}

\usepackage{cite}
\usepackage{amsmath,amssymb,amsfonts}
\usepackage{graphicx}
\usepackage{booktabs}
\usepackage{multirow}
\usepackage{array}
\usepackage{url}
\usepackage{hyperref}
\usepackage{enumitem}
\usepackage{algorithm}
\usepackage{algorithmic}
\usepackage{tikz}
\usetikzlibrary{arrows.meta,positioning,shapes.geometric,fit,backgrounds,calc}

\begin{document}

\title{
RISC-V Functional Safety for Autonomous Automotive Systems:\\
An Analytical Framework and Research Roadmap for\\ML-Assisted Certification
}

\author{
\IEEEauthorblockN{N.~Andreasyan\IEEEauthorrefmark{1},
M.~Struve\IEEEauthorrefmark{1},
A.~Popov\IEEEauthorrefmark{2},
M.~Nikolaev\IEEEauthorrefmark{2},
and V.~Vashkelis\IEEEauthorrefmark{2}}
\IEEEauthorblockA{
\IEEEauthorrefmark{1}Automotive Safety Lab\\
\IEEEauthorrefmark{2}Embedded Intelligence Lab\\
\texttt{\{research,safety\}@emilab.org}
}
}

\maketitle

\begin{abstract}

RISC-V is emerging as a viable platform for automotive-grade
embedded computing, with recent ISO~26262 ASIL-D certifications
of commercial RISC-V processor IP demonstrating readiness for
safety-critical deployment in autonomous driving systems.
However, functional safety in automotive systems is
fundamentally a certification problem rather than a processor
problem: the dominant costs arise from diagnostic coverage
analysis, toolchain qualification, fault injection campaigns,
safety-case generation, and compliance with ISO~26262,
ISO~21448 (SOTIF), and ISO/SAE~21434.

This paper provides a structured analytical framework and
expert-driven research roadmap for enabling economically
superior certification of automotive-grade RISC-V platforms.
Rather than proposing a single algorithmic breakthrough, we
contribute: (1)~a \emph{certification economics} framework
with formal cost and advantage models for strategic decision
support; (2)~an \emph{ML-assisted certification framework}
mapping LLMs, knowledge graphs, reinforcement learning, and
graph neural networks to specific certification bottlenecks;
(3)~a five-level \emph{RISC-V Safety Maturity Model} (RSMM)
for evaluating platform certifiability; and (4)~an
illustrative case study applying the framework to an ASIL-D
autonomous emergency braking (AEB) ECU.

We present a structured comparison of Arm and RISC-V
architectures across eight functional safety dimensions,
address the convergence of ISO~26262, ISO~21448, and
ISO/SAE~21434 for autonomous driving, and define quantitative
evaluation metrics for certification cost reduction. The paper
is positioned as an analytical perspective contribution that
structures the research space and identifies the highest-value
problems for the RISC-V automotive certification ecosystem.

\end{abstract}


\section{Introduction}
\label{sec:introduction}

The automotive industry is undergoing a structural
transformation from distributed Electronic Control Units
(ECUs) toward centralized compute platforms and
Software-Defined Vehicles (SDVs). Advanced Driver Assistance
Systems (ADAS), Level-3 and Level-4 autonomous driving,
battery management, zonal architectures, and over-the-air
(OTA) software updates are fundamentally changing the safety
requirements of automotive systems
\cite{autonomous2024,bmw2025,koopman2019}.

Functional safety has consequently become a first-order
engineering constraint. ISO~26262 defines the framework
for functional safety of electrical and electronic systems
in road vehicles through Automotive Safety Integrity Levels
(ASIL A--D), where ASIL-D represents the highest safety
criticality \cite{iso26262}. For autonomous driving, ISO~21448
(SOTIF) addresses safety of the intended functionality,
covering perception failures and unknown unsafe scenarios
that fall outside the scope of ISO~26262 \cite{sotif2022}.
Simultaneously, ISO/SAE~21434 introduces cybersecurity
engineering requirements that are now inseparable from
functional safety \cite{iso21434}.

Historically, safety-critical automotive systems have relied
on proprietary architectures, particularly Arm-based lockstep
microcontrollers and DSP-centric controllers
\cite{arm2023,infineon2023}. RISC-V introduces a fundamentally
different paradigm: open ISA governance, transparent
extensibility, and vendor independence
\cite{automotiveecu2023,riscvisa2019}. This transition is
no longer theoretical. In 2025, Andes Technology announced
that its D45-SE processor achieved full ISO~26262 ASIL-D
certification by SGS-T\"{U}V Saar, integrating dual-core
lockstep (DCLS), ECC protection, bus protection, stack
protection, and real-time diagnostic safety circuits
\cite{andes2025}. Similar developments are visible across
the RISC-V automotive ecosystem
\cite{andes2022,d23se2025,sifive2023,codasip2024}.

This changes the central research question from
\emph{``Can RISC-V be used in automotive systems?''} to:
\textbf{How can RISC-V become the preferred certification
platform for autonomous driving systems?}

This paper addresses that question by providing a structured
analytical framework, a maturity model, and an ML-assisted
certification methodology specifically designed for
automotive RISC-V platforms.

\smallskip
\noindent\textbf{Scope of This Work.}
This paper is intended as an analytical perspective paper
rather than an experimental implementation study. It does not
propose a single isolated algorithmic breakthrough. Instead,
it provides a structured analytical framework and
expert-driven roadmap for enabling economically superior
certification of automotive-grade RISC-V platforms for
autonomous driving systems. Its purpose is to structure the
research space, identify the highest-value problems, and
define a formal basis for certification-oriented RISC-V
development. Experimental validation of individual framework
components is identified as future work.

\section{Contributions}
\label{sec:contributions}

This paper makes the following strategic framework
contributions:

\begin{enumerate}[leftmargin=*]
\item \textbf{Certification Economics Framework.}
We introduce the concept of certification economics
for automotive-grade RISC-V systems and formalize it
through an analytical cost model, an ML optimization
gain model, and a RISC-V Certification Advantage (RCA)
score. These are presented as strategic decision-support
models for evaluating certification investments, not as
experimentally validated predictive models.

\item \textbf{ML-Assisted Certification Framework.}
We propose a systematic mapping of ML methods---specifically
LLMs with knowledge graphs, reinforcement learning, and
graph neural networks---to RISC-V certification bottlenecks.
For each primary method, we provide a structured justification
of selection over alternatives. ML is treated as an enabler
of certification strategy, not as the central research topic.

\item \textbf{RISC-V Safety Maturity Model (RSMM).}
We define a five-level maturity assessment framework for
evaluating the certifiability of automotive RISC-V platforms,
intended to help researchers prioritize work, vendors assess
readiness, OEMs evaluate certifiability, and assessment
bodies structure expectations.

\item \textbf{Illustrative Case Study.}
We apply the framework to an ASIL-D Autonomous Emergency
Braking (AEB) ECU scenario, demonstrating where certification
cost appears, where RISC-V provides structural advantages,
where ML assistance is most impactful, and how the RSMM
applies in practice.
\end{enumerate}

\section{Related Work}
\label{sec:related}

\subsection{Automotive RISC-V Safety}

The emergence of RISC-V in automotive applications has been
documented by several recent efforts. Cuomo et al.\
\cite{automotiveecu2023} present an open RISC-V platform
for next-generation automotive ECUs. Andes Technology has
achieved both ASIL-D development process certification
\cite{andes2022} and ASIL-D product certification for the
D45-SE \cite{andes2025} and D23-SE \cite{d23se2025} cores.
SiFive \cite{sifive2023} and Codasip \cite{codasip2024} have
introduced automotive-grade RISC-V IP with ISO~26262
compliance targets. Pinto et al.\ \cite{securewheels2024}
identify security gaps in RISC-V MCU architectures,
particularly regarding initiator-side protection for
mixed-criticality SDV architectures.

\subsection{ML for Safety Certification}

The application of ML to safety engineering is an emerging
field. Salay et al.\ \cite{salay2017,salay2018} analyze
the gaps between ISO~26262 and ML-specific lifecycle
requirements. Iyenghar et al.\ \cite{iyenghar2024} propose
systematic enhancements to ISO~26262 with ML-specific
testing methods. Cheng et al.\ \cite{cheng2020} provide
a quantitative projection of ISO~26262 requirements onto
ML-based functions. Kochanthara et al.\
\cite{kochanthara2021} present safety case patterns for
systems with ML components.

\subsection{ML Safety in Autonomous Driving}

Vyas and Xu \cite{autonomous2024} provide an overview of
safety design challenges in AI-driven autonomous vehicles.
Werling et al.\ \cite{bmw2025} present a safety integrity
framework for automated driving. The SMIRK project
\cite{smirk2022} demonstrates a complete safety case for
an ML component in a pedestrian AEB system. Burton et al.\
\cite{safecomp2017} address safety argumentation for ML
in highly automated driving. Koopman and Wagner
\cite{koopman2019} frame autonomous vehicle safety as an
interdisciplinary challenge spanning engineering, policy,
and validation methodology.

\subsection{Formal Verification for RISC-V}

Reid \cite{reid2016} describes industrial-scale ISA formal
verification at Arm, providing a benchmark for RISC-V
efforts. Wolf et al.\ \cite{riscvformal2018} present
open-source formal verification frameworks for RISC-V cores.
These approaches are directly relevant to ASIL-D
certification, where formal correctness proofs can
substantially reduce diagnostic coverage requirements
\cite{iec61508}.

\subsection{Standards Convergence: ISO~26262, ISO~21448, and ISO/SAE~21434}

Recent work has highlighted the need to address functional
safety, safety of the intended functionality, and
cybersecurity as an integrated concern for autonomous
driving. Schildbach \cite{schildbach2018} examines ISO~26262
application in automated vehicle control. Bastos et al.\
\cite{bridge2025} provide a comparative analysis of
ISO~21434, ISO~26262, and ML requirements. The SOTIF
standard (ISO~21448) \cite{sotif2022} addresses perception
and decision-making failures that are not covered by
ISO~26262's random hardware fault model, making it essential
for autonomous driving systems. The split-and-cover
methodology \cite{splitcover2025} demonstrates SystemC-based
FMEDA improvement under ISO~26262. Macher et al.\
\cite{fmeda2019} document FMEDA challenges and best
practices in automotive contexts.

\section{Why RISC-V Is Structurally Different for Automotive Safety}
\label{sec:riscv_advantages}

RISC-V should not be analyzed as merely another processor
ISA. Its architectural properties create structural
advantages for functional safety engineering that are
qualitatively different from proprietary alternatives.

\subsection{Open ISA and Certification Transparency}

Unlike proprietary architectures, RISC-V specifications are
fully open and auditable \cite{riscvisa2019}. This
transparency directly improves traceability from safety
requirements to implementation, formal verification
feasibility, toolchain qualification transparency, and
long-term certification maintainability. In contrast,
Arm-based platforms restrict ISA-level access through NDA
agreements, limiting independent verification and creating
structural opacity in the certification evidence chain
\cite{arm2023,reid2016}.

\subsection{Custom Extensions with Controlled Safety Scope}

Automotive systems increasingly require domain-specific
extensions for motor control, cryptographic operations,
sensor fusion, deterministic real-time execution, and
autonomous driving acceleration. RISC-V enables controlled
ISA customization through its standard extension mechanism
\cite{riscvisa2019}. However, this introduces a critical
certification challenge: \emph{customization must not
invalidate portability or existing certification evidence}.
This tension between extensibility and certification
stability requires careful architectural governance and
represents a uniquely RISC-V research problem.

\subsection{Formal Verification Opportunity}

Because the ISA specification is open, formal verification
from ISA semantics to microarchitectural implementation
becomes significantly more practical than in closed
ecosystems \cite{riscvformal2018}. This is especially
valuable for verifying PMP/MMU correctness, privilege
isolation, lockstep equivalence, debug access control,
and fault containment boundaries---all of which are
critical for ASIL-D compliance.

\subsection{Toolchain Qualification and Certification Reuse}
\label{sec:toolchain}

Toolchain qualification represents one of the most
significant and least discussed certification cost drivers.
Without a qualified compiler, ASIL certification becomes
substantially more expensive due to increased verification
requirements \cite{iso26262,hightec2024}.

RISC-V creates a unique opportunity for
\emph{qualification-by-construction} approaches to compiler
certification. The open nature of both the ISA and
open-source toolchains (LLVM, GCC) enables: (i)~deterministic
compilation with traceable optimization passes,
(ii)~compiler confidence level assessment per ISO~26262
Tool Confidence Level (TCL) requirements,
(iii)~certification evidence reuse across RISC-V
implementations sharing the same ISA subset, and
(iv)~community-driven qualification artifacts that reduce
per-vendor qualification cost \cite{hightec2024}.

This is in contrast to proprietary architectures, where
toolchain qualification is vendor-locked and
non-transferable across implementations.

\subsection{Security and Functional Safety Convergence}

Pinto et al.\ \cite{securewheels2024} demonstrate that
automotive RISC-V systems require stronger security
primitives for virtualized MCUs and initiator-side
protection, especially for ISO/SAE~21434 compliance in
mixed-criticality SDV architectures. This convergence of
safety and security is particularly important for autonomous
driving platforms where attack surfaces expand with
connectivity \cite{bosch2021}.

\subsection{Structured Comparison: Arm vs.\ RISC-V}

Table~\ref{tab:arm_vs_riscv} presents a systematic
comparison across eight functional safety dimensions.

\begin{table}[ht]
\caption{Arm vs.\ RISC-V for Functional Safety Certification}
\centering
\begin{tabular}{p{4.0cm} p{3.5cm} p{3.5cm}}
\toprule
\textbf{Dimension} & \textbf{Arm} & \textbf{RISC-V} \\
\midrule
ISA Transparency
& Proprietary, NDA-restricted
& Fully open and auditable \\
Formal Verification
& Limited by IP access
& Full RTL verification feasible \\
Custom Extensions
& Vendor-controlled (CIE)
& Open standard mechanism \\
Toolchain Qualification
& Proprietary, vendor-locked
& Open-source + commercial options \\
Certification Reuse
& Vendor-dependent licensing
& Portable across implementations \\
Vendor Lock-in
& High (IP licensing model)
& Low (open specification) \\
Debug Transparency
& Proprietary debug IP
& Open debug specification \\
Supply-Chain Independence
& Single-vendor dependency
& Multi-vendor ecosystem \\
\bottomrule
\end{tabular}
\label{tab:arm_vs_riscv}
\end{table}

\section{Safety Requirements of Autonomous Driving Systems}
\label{sec:ad_safety}

Autonomous driving systems introduce fundamentally different
safety requirements compared to classical ECUs. The challenge
is no longer protecting a single controller but ensuring
system-wide safety across the entire autonomous driving stack
\cite{autonomous2024,bmw2025,koopman2019}.

\subsection{Standards Landscape for Autonomous Driving}

Autonomous driving systems operate at the intersection of
three complementary standards:

\begin{itemize}[leftmargin=*]
\item \textbf{ISO~26262} addresses random hardware faults
and systematic software failures through ASIL classification
and diagnostic coverage requirements \cite{iso26262}.

\item \textbf{ISO~21448 (SOTIF)} addresses safety of the
intended functionality---specifically, hazardous behavior
caused by functional insufficiencies or foreseeable misuse,
including perception uncertainty, unknown unsafe scenarios,
sensor ambiguity, planning uncertainty, and ML model
uncertainty \cite{sotif2022}.

\item \textbf{ISO/SAE~21434} addresses cybersecurity
engineering, including threat analysis and risk assessment
(TARA), which is inseparable from functional safety in
connected autonomous vehicles \cite{iso21434}.
\end{itemize}

The convergence of these three standards creates a
certification challenge that is qualitatively more complex
than traditional ECU certification. RISC-V platforms for
autonomous driving must address all three simultaneously.

\subsection{Autonomous Driving Safety Stack and RISC-V Implications}

Table~\ref{tab:ad_stack} maps the autonomous driving safety
stack to specific RISC-V architectural implications.

\begin{table}[ht]
\caption{Autonomous Driving Safety Stack and RISC-V Implications}
\centering
\begin{tabular}{p{2.8cm} p{3.8cm} p{4.2cm}}
\toprule
\textbf{AD Layer} & \textbf{Safety Requirement} &
\textbf{RISC-V Implication} \\
\midrule
Perception &
Bounded latency, degraded-safe output under sensor failure &
Deterministic execution, custom sensor fusion extensions \\
Planning &
Deterministic execution, uncertainty handling &
Formal verifiability, WCET guarantees \\
Actuation &
ASIL-D integrity, sub-ms response &
DCLS, safety islands, hardware interlocks \\
Sensor Fusion &
Heterogeneous input (radar, lidar, camera) &
Multi-modal I/O, DMA protection, PMP isolation \\
Redundancy &
Fail-operational behavior &
Split-lock architecture, independent safety cores \\
Zonal Control &
Distributed safety across domains &
Adaptive lockstep, inter-zone isolation \\
Degraded Mode &
Graceful performance reduction &
Safe-state transition, watchdog supervision \\
\bottomrule
\end{tabular}
\label{tab:ad_stack}
\end{table}

\subsection{Mixed-Criticality Compute}

ADAS and autonomous driving combine ASIL-D braking and
steering, ASIL-B perception pipelines, QM infotainment
workloads, and Linux/RTOS coexistence on shared hardware.
This requires strict temporal and spatial isolation with
predictable scheduling guarantees
\cite{automotiveecu2023,securewheels2024}.

\subsection{Continuous Certification}

Unlike traditional ECUs, autonomous vehicles are updated
continuously via OTA deployment. This creates recurring
certification challenges: re-validation, regression
detection, compiler re-qualification, and evidence
regeneration after each update cycle
\cite{bridge2025,iyenghar2024,nhtsa2022}.

\section{Proposed Certification Economics Framework}
\label{sec:cert_economics}

A common misconception is that automotive functional safety
is primarily a hardware problem. In practice, the dominant
cost drivers are engineering activities: FMEDA/FMEA
generation, diagnostic coverage analysis, toolchain
qualification, safety case documentation, assessment body
audit cycles, evidence traceability, fault injection
campaigns, and re-certification after OTA updates
\cite{fmeda2019}.

We define \emph{certification economics} as the optimization
of engineering effort, traceability, and compliance cost
required to achieve and maintain ASIL certification.

The models presented below are intended as \emph{analytical
decision-support tools} for strategic evaluation of
certification investments. They provide a structured basis
for comparing architectural alternatives and prioritizing
automation efforts. Industrial benchmarking and empirical
calibration of model parameters are identified as essential
future work.

\subsection{Certification Cost Function}

We model the total certification cost as:
\begin{equation}
C_{\text{total}} = C_{\text{FMEDA}} + C_{\text{verif}}
  + C_{\text{tool}} + C_{\text{FI}}
  + C_{\text{audit}} + C_{\text{recert}}
\label{eq:cost}
\end{equation}
where $C_{\text{FMEDA}}$ represents FMEDA and FMEA
generation cost, $C_{\text{verif}}$ captures formal and
simulation-based verification effort, $C_{\text{tool}}$
denotes toolchain qualification cost, $C_{\text{FI}}$
represents fault injection campaign cost,
$C_{\text{audit}}$ captures assessment body audit effort,
and $C_{\text{recert}}$ represents the recurring cost of
re-certification after OTA updates.

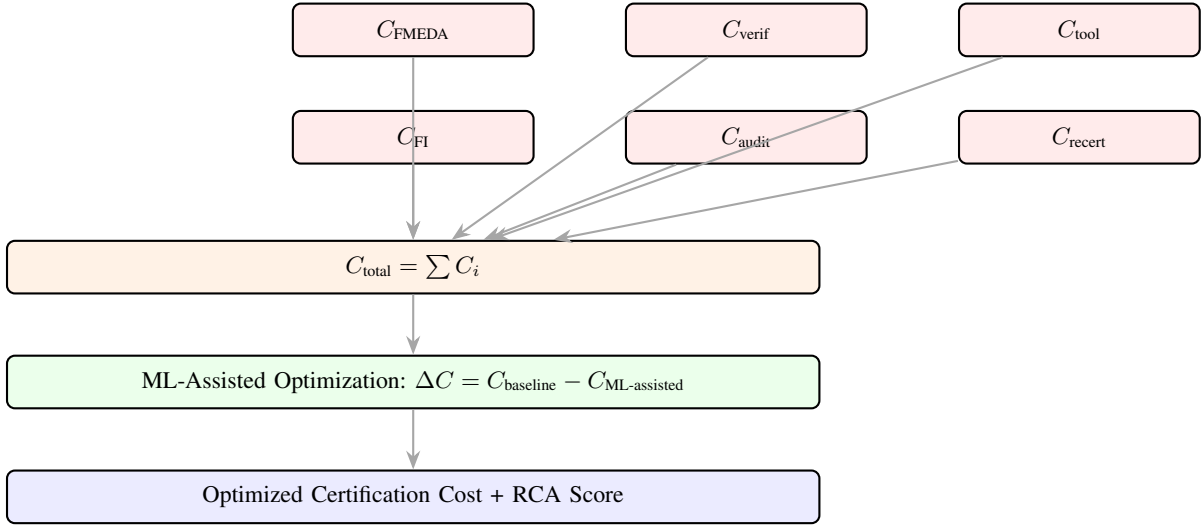
\begin{figure}[ht]
\centering
\begin{tikzpicture}[
  node distance=0.7cm and 1.2cm,
  box/.style={draw, rounded corners=3pt, minimum width=3.2cm,
    minimum height=0.7cm, align=center, font=\small, thick},
  arrow/.style={-{Stealth[length=2.5mm]}, thick, gray!70},
  costbox/.style={box, fill=red!8},
  mlbox/.style={box, fill=green!8},
  outbox/.style={box, fill=blue!8}
]
\node[costbox] (fmeda) {$C_{\text{FMEDA}}$};
\node[costbox, right=of fmeda] (verif) {$C_{\text{verif}}$};
\node[costbox, right=of verif] (tool) {$C_{\text{tool}}$};
\node[costbox, below=of fmeda] (fi) {$C_{\text{FI}}$};
\node[costbox, right=of fi] (audit) {$C_{\text{audit}}$};
\node[costbox, right=of audit] (recert) {$C_{\text{recert}}$};
\node[box, fill=orange!10, below=1.0cm of fi,
  minimum width=10.8cm] (total)
  {$C_{\text{total}} = \sum C_i$};
\node[mlbox, below=0.8cm of total,
  minimum width=10.8cm] (ml)
  {ML-Assisted Optimization: $\Delta C = C_{\text{baseline}} - C_{\text{ML-assisted}}$};
\node[outbox, below=0.8cm of ml,
  minimum width=10.8cm] (out)
  {Optimized Certification Cost + RCA Score};
\draw[arrow] (fmeda) -- (total);
\draw[arrow] (verif) -- (total);
\draw[arrow] (tool) -- (total);
\draw[arrow] (fi) -- (total);
\draw[arrow] (audit) -- (total);
\draw[arrow] (recert) -- (total);
\draw[arrow] (total) -- (ml);
\draw[arrow] (ml) -- (out);
\end{tikzpicture}
\caption{Certification economics optimization flow:
cost decomposition, ML-assisted reduction, and
advantage scoring.}
\label{fig:cert_economics}
\end{figure}

Figure~\ref{fig:cert_economics} illustrates the
certification economics optimization flow.

\subsection{ML Optimization Gain}

The certification cost reduction achievable through
ML-assisted automation is modeled as:
\begin{equation}
\Delta C = C_{\text{baseline}} - C_{\text{ML-assisted}}
\label{eq:gain}
\end{equation}
where $C_{\text{baseline}}$ is the cost under traditional
manual certification workflows and $C_{\text{ML-assisted}}$
is the cost with ML-augmented processes. The optimization
objective is to maximize $\Delta C$ while maintaining
certification validity and audit confidence.

\subsection{RISC-V Certification Advantage Score}

We define the RISC-V Certification Advantage (RCA) score
as a weighted composite metric:
\begin{equation}
RCA = \alpha \cdot T + \beta \cdot V
    + \gamma \cdot F + \delta \cdot Q
\label{eq:rca}
\end{equation}
where $T$ represents ISA transparency (auditable
specification access), $V$ denotes formal verifiability
(feasibility of ISA-to-RTL proofs), $F$ captures formal
proof feasibility (coverage of safety properties), and $Q$
quantifies qualification cost reduction (toolchain and
evidence reuse). The coefficients $\alpha, \beta, \gamma,
\delta \in [0,1]$ with $\alpha + \beta + \gamma + \delta = 1$
are context-dependent and determined by the specific ASIL
target and certification scope.

The RCA score is a structured evaluation tool, not a
predictive formula. Its value lies in enabling systematic
comparison of architectural alternatives for certification
cost optimization.

\section{ML-Assisted Certification Framework}
\label{sec:ml_framework}

Machine Learning is not the central topic of this paper.
Rather, ML is treated as an enabling technology that addresses
specific bottlenecks in RISC-V certification workflows. The
guiding question is: \emph{How can ML reduce the cost and
uncertainty of certifying automotive-grade RISC-V systems?}

We focus on three primary ML integration points where the
strongest certification impact is expected, with structured
justification for each.

\subsection{LLM + Knowledge Graph for FMEDA and Safety Case Automation}

\textbf{Problem.} FMEDA generation remains largely manual,
spreadsheet-driven, and expensive \cite{fmeda2019,splitcover2025}.
Safety documentation requires maintaining traceable links
from hazard analysis (HARA) through FMEA, FMEDA, verification
evidence, and compliance mapping to specific ISO~26262 clauses.

\textbf{Why LLMs + Knowledge Graphs.} Large Language Models
can process heterogeneous technical documents (RTL
specifications, safety manuals, requirements databases) and
generate structured outputs including preliminary FMEDA
drafts, failure mode extractions, and requirement-to-failure
traceability mappings \cite{mdpi2024}. However, LLMs alone
suffer from hallucination risk that is unacceptable in
safety-critical certification contexts. We therefore propose
a combined architecture: a certification knowledge graph
provides the structural backbone encoding the traceability
chain
$Requirement \rightarrow Implementation \rightarrow
Verification \rightarrow Evidence \rightarrow ISO\ Clause$,
while LLM-based agents perform document parsing and draft
generation with graph-grounded retrieval to mitigate
hallucination \cite{safecomp2017,kochanthara2021}.

\textbf{Why not alternatives.} Rule-based systems lack
flexibility across diverse RISC-V implementations.
Traditional NLP approaches (BERT, Word2Vec) support specific
subtasks \cite{mdpi2024} but lack generative capability.
LLMs without knowledge graph grounding introduce
unacceptable hallucination risk for certification evidence.

\subsection{Reinforcement Learning for Fault Injection Optimization}

\textbf{Problem.} ISO~26262 requires extensive fault
injection testing. Brute-force campaigns across the full
fault space are prohibitively expensive for complex
automotive SoCs \cite{iso26262,fmeda2019}.

\textbf{Why RL.} Fault injection optimization is a sequential
decision problem: at each step, the agent selects the next
fault to inject based on observed coverage improvement. This
maps naturally to a Markov Decision Process where the state
represents current diagnostic coverage, actions are fault
selections, and the reward is coverage gain per injection cost.

\textbf{Why not supervised learning.} There are no fixed
ground-truth labels for optimal injection sequences.
\textbf{Why not static heuristics.} Static strategies
provide poor exploration of complex fault spaces and cannot
adapt to implementation-specific coverage distributions.

\subsection{GNNs for Diagnostic Coverage Optimization}

\textbf{Problem.} Diagnostic coverage is often the deciding
factor for ASIL compliance. Optimizing safety mechanism
placement requires understanding fault propagation across
hierarchical hardware structures \cite{splitcover2025}.

\textbf{Why GNNs.} The hardware architecture of an automotive
SoC is naturally represented as a graph. Graph Neural Networks
can model fault propagation paths, predict latent fault
manifestation, optimize safety monitor placement, and determine
optimal supervision boundaries for lockstep systems, safety
islands, and mixed-criticality SoCs.

\textbf{Why not traditional analysis.} Manual fault
propagation analysis does not scale to modern SoC complexity.
Statistical methods lack the structural awareness that
graph-based approaches provide.

\subsection{ML Method Summary}

Table~\ref{tab:framework} summarizes the primary and
secondary ML integration points.

\begin{table}[ht]
\caption{ML-Assisted Certification: Method Mapping and Priority}
\centering
\begin{tabular}{p{4.2cm} p{3.6cm} p{1.8cm}}
\toprule
\textbf{Certification Problem} &
\textbf{ML Method} &
\textbf{Priority} \\
\midrule
FMEDA generation &
LLM + NLP &
Very High \\
Safety case automation &
Knowledge Graph + LLM &
Very High \\
Fault injection optimization &
Reinforcement Learning &
Very High \\
Diagnostic coverage analysis &
Graph Neural Networks &
Very High \\
Compiler qualification &
Anomaly Detection &
High \\
\bottomrule
\end{tabular}
\label{tab:framework}
\end{table}

\subsection{Trustworthiness and Limitations of ML in Certification}
\label{sec:ml_limits}

The application of ML to safety certification introduces
inherent trustworthiness challenges that must be explicitly
acknowledged:

\begin{itemize}[leftmargin=*]
\item \textbf{LLM hallucination risk.} Generated FMEDA
entries or safety case fragments may contain plausible but
factually incorrect content. All ML-generated certification
artifacts require expert review and formal validation before
inclusion in audit evidence.

\item \textbf{Certification authority acceptance.} Assessment
bodies (e.g., T\"{U}V, UL) have not yet established formal
acceptance criteria for ML-generated certification evidence.
Regulatory framework development is a prerequisite for
production deployment of ML-assisted certification
\cite{ul2023}.

\item \textbf{Validation of ML-generated evidence.}
ML outputs used in certification must themselves be
validated to a confidence level commensurate with the
target ASIL. This creates a meta-certification challenge.

\item \textbf{OEM trust constraints.} Automotive OEMs
require deterministic, auditable processes. ML-assisted
workflows must provide full traceability and
explainability to meet OEM governance requirements.

\item \textbf{Proprietary data limitations.} Training
ML models for certification tasks requires access to
FMEDA databases, safety cases, and fault injection
results that are typically proprietary and commercially
sensitive.
\end{itemize}

These limitations do not invalidate the proposed framework
but define the boundaries within which ML-assisted
certification can be responsibly deployed.

\section{RISC-V Safety Maturity Model}
\label{sec:maturity}

To provide a structured evaluation framework for automotive
RISC-V platforms, we propose a five-level \emph{RISC-V
Safety Maturity Model} (RSMM). The RSMM is designed as a
\emph{maturity assessment tool} rather than a strict
quantitative instrument. Its purpose is to enable:

\begin{itemize}[leftmargin=*]
\item \textbf{Researchers:} prioritize work toward the
highest-impact certification gaps.
\item \textbf{Vendors:} assess readiness of their RISC-V
IP for specific ASIL targets.
\item \textbf{OEMs:} evaluate certifiability of candidate
platforms for vehicle programs.
\item \textbf{Assessment bodies:} structure certification
expectations for RISC-V-based systems.
\end{itemize}

Table~\ref{tab:maturity} defines the five levels.

\begin{table}[ht]
\caption{RISC-V Safety Maturity Model (RSMM)}
\centering
\begin{tabular}{c p{3.6cm} p{6.5cm}}
\toprule
\textbf{Level} & \textbf{Classification} & \textbf{Characteristics} \\
\midrule
1 & Safety-Aware MCU &
Basic safety features (ECC, watchdog), no formal ASIL
target, suitable for QM applications. \\
2 & ASIL-B Capable System &
Hardware fault metrics meet ASIL-B, partial FMEDA, basic
diagnostic coverage, initial toolchain qualification. \\
3 & ASIL-D Lockstep Platform &
Full DCLS, comprehensive FMEDA, diagnostic coverage
$\geq 99\%$, qualified toolchain, T\"{U}V-assessed
safety manual. \\
4 & Mixed-Criticality AD Platform &
Hypervisor-certified isolation, ASIL-D + ASIL-B + QM
coexistence, safety islands, deterministic scheduling,
SOTIF-aware design. \\
5 & Continuous-Certification OTA Platform &
Incremental re-certification, ML-assisted evidence
regeneration, regression-aware validation, live safety
case updates. \\
\bottomrule
\end{tabular}
\label{tab:maturity}
\end{table}

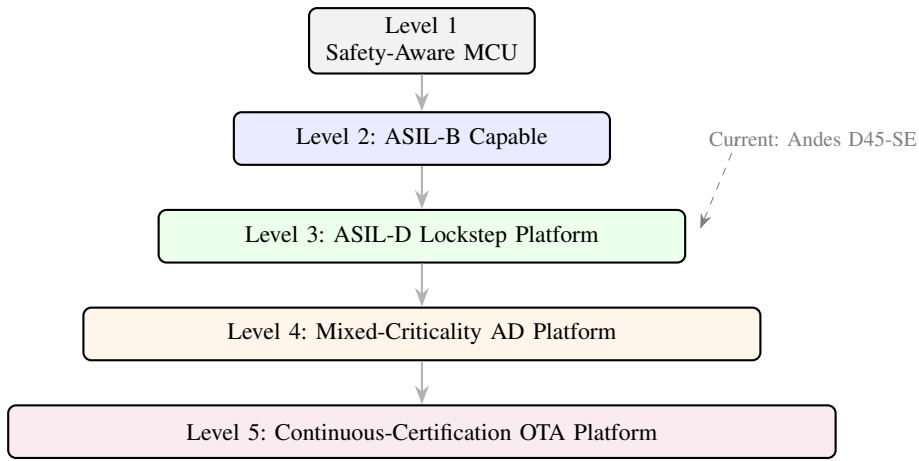
\begin{figure}[ht]
\centering
\begin{tikzpicture}[
  level/.style={draw, thick, rounded corners=3pt,
    minimum height=0.7cm, align=center, font=\small},
  arrow/.style={-{Stealth[length=2.5mm]}, thick, gray!60}
]
\node[level, fill=gray!10, minimum width=3cm] (l1) at (0,0)
  {Level 1\\Safety-Aware MCU};
\node[level, fill=blue!8, minimum width=5cm] (l2) at (0,-1.3)
  {Level 2: ASIL-B Capable};
\node[level, fill=green!8, minimum width=7cm] (l3) at (0,-2.6)
  {Level 3: ASIL-D Lockstep Platform};
\node[level, fill=orange!8, minimum width=9cm] (l4) at (0,-3.9)
  {Level 4: Mixed-Criticality AD Platform};
\node[level, fill=purple!8, minimum width=11cm] (l5) at (0,-5.2)
  {Level 5: Continuous-Certification OTA Platform};
\draw[arrow] (l1) -- (l2);
\draw[arrow] (l2) -- (l3);
\draw[arrow] (l3) -- (l4);
\draw[arrow] (l4) -- (l5);
\node[font=\footnotesize, gray] at (5.2,-1.3)
  {Current: Andes D45-SE};
\draw[-{Stealth[length=2mm]}, dashed, gray]
  (4.2,-1.3) -- (3.7,-2.5);
\end{tikzpicture}
\caption{RISC-V Safety Maturity Model progression.
Current commercially certified platforms correspond
to Level~3.}
\label{fig:rsmm}
\end{figure}

Figure~\ref{fig:rsmm} illustrates the RSMM progression.

Current commercially certified platforms such as the
Andes D45-SE correspond to Level~3 \cite{andes2025}. The
transition from Level~3 to Level~5 represents the primary
research and engineering challenge for the RISC-V automotive
ecosystem.

\section{RISC-V Safety Architecture Patterns}
\label{sec:architecture}

Three dominant architectural patterns emerge for
ASIL-D-capable RISC-V platforms.

\subsection{Dual-Core Lockstep (DCLS)}

DCLS remains the dominant ASIL-D safety mechanism, in which
two identical cores execute the same instruction stream while
comparator logic detects divergence. The Andes D45-SE
integrates DCLS with real-time diagnostic circuits
\cite{andes2025}. Open research directions include delayed
lockstep, split-lock architectures \cite{d23se2025},
heterogeneous lockstep, and adaptive lockstep for zonal
controllers.

\subsection{Safety Islands}

Dedicated safety cores supervise watchdog functions, emergency
state transitions, fault detection, and safety policy
enforcement. Safety islands are critical for centralized
autonomous platforms where independent supervision of the
main compute cluster is required \cite{infineon2023}.

\subsection{Mixed-Criticality Isolation}

Automotive RISC-V platforms require hypervisor-certified
separation for Linux and AUTOSAR coexistence, deterministic
interrupt behavior, and memory protection through PMP and
hardware partitioning \cite{securewheels2024}. This
represents a system-level challenge rather than a core-level
problem.

\section{Illustrative Case Study: ASIL-D AEB ECU}
\label{sec:case_study}

To demonstrate practical application of the proposed
framework, we present an illustrative scenario based on an
ASIL-D Autonomous Emergency Braking (AEB) ECU. This is not
an experimental validation but an \emph{application example}
that grounds the analytical framework in a realistic
certification context.

\subsection{System Description}

Consider an AEB ECU based on a RISC-V DCLS platform
(RSMM Level~3) responsible for: (i)~radar and camera
sensor fusion, (ii)~collision risk assessment,
(iii)~emergency braking actuation at ASIL-D integrity,
and (iv)~fail-safe degradation under sensor failure.

\subsection{Certification Cost Analysis}

Applying the cost model (Eq.~\ref{eq:cost}):

\begin{itemize}[leftmargin=*]
\item $C_{\text{FMEDA}}$: The AEB ECU requires comprehensive
FMEDA covering the DCLS core, comparator logic, ECC memory,
bus protection, and safety island. This is the highest
single cost component, typically requiring 6--12 months of
expert engineering effort.

\item $C_{\text{tool}}$: Compiler qualification for the
RISC-V toolchain (GCC or LLVM) at TCL-2 or TCL-3 is
required. RISC-V's open toolchain enables qualification
evidence reuse, reducing this cost compared to proprietary
alternatives (Section~\ref{sec:toolchain}).

\item $C_{\text{FI}}$: Fault injection campaigns must cover
the full DCLS fault space. RL-based optimization
(Section~\ref{sec:ml_framework}) targets this cost component.

\item $C_{\text{audit}}$: T\"{U}V assessment requires
complete traceability from requirements to evidence.
Knowledge-graph-based automation directly reduces audit
preparation cost.

\item $C_{\text{recert}}$: OTA firmware updates to the AEB
ECU trigger re-certification requirements, making this a
recurring cost that grows with vehicle lifetime.
\end{itemize}

\subsection{RISC-V Advantage Assessment}

For this AEB scenario, the RCA score (Eq.~\ref{eq:rca})
is dominated by:
\begin{itemize}[leftmargin=*]
\item High $T$ (transparency): open ISA enables independent
FMEDA verification.
\item High $V$ (verifiability): DCLS equivalence is formally
provable.
\item High $Q$ (qualification reuse): compiler qualification
evidence is portable across RISC-V implementations.
\end{itemize}

\subsection{RSMM Application}

The AEB ECU requires RSMM Level~3 (ASIL-D Lockstep Platform)
as a minimum. If the ECU must support OTA-updateable perception
models, it approaches Level~4 (Mixed-Criticality AD Platform)
requirements. The RSMM provides a structured basis for
evaluating whether a candidate RISC-V platform meets the
certification readiness requirements for this application.

\section{Evaluation Methodology}
\label{sec:evaluation}

We define the evaluation metrics that would be used to
assess each component of the framework in future industrial
deployment.

\subsection{Certification Cost Metrics}

\begin{itemize}[leftmargin=*]
\item \textbf{FMEDA effort reduction (\%):}
Reduction in person-hours for FMEDA generation using
LLM-assisted automation compared to manual baseline.

\item \textbf{Certification cycle time reduction:}
Reduction in calendar time from project start to
T\"{U}V assessment completion.

\item \textbf{Fault injection efficiency:}
Diagnostic coverage achieved per unit of fault injection
campaign cost (coverage-per-dollar).

\item \textbf{Audit effort reduction:}
Reduction in assessor review hours attributable to
improved evidence traceability.

\item \textbf{Re-certification latency:}
Time from OTA update deployment to updated certification
evidence availability.
\end{itemize}

\subsection{ML Component Metrics}

\begin{itemize}[leftmargin=*]
\item \textbf{Traceability completeness:}
Percentage of requirements with verified bidirectional links
to implementation, verification, and evidence.

\item \textbf{False positive rate:}
Rate of incorrect faults or spurious safety mechanism
recommendations by ML components.

\item \textbf{Safety case coverage:}
Percentage of ISO~26262 clauses with automatically generated
evidence linkage.
\end{itemize}

\section{Proposed Certification Workflow}
\label{sec:workflow}

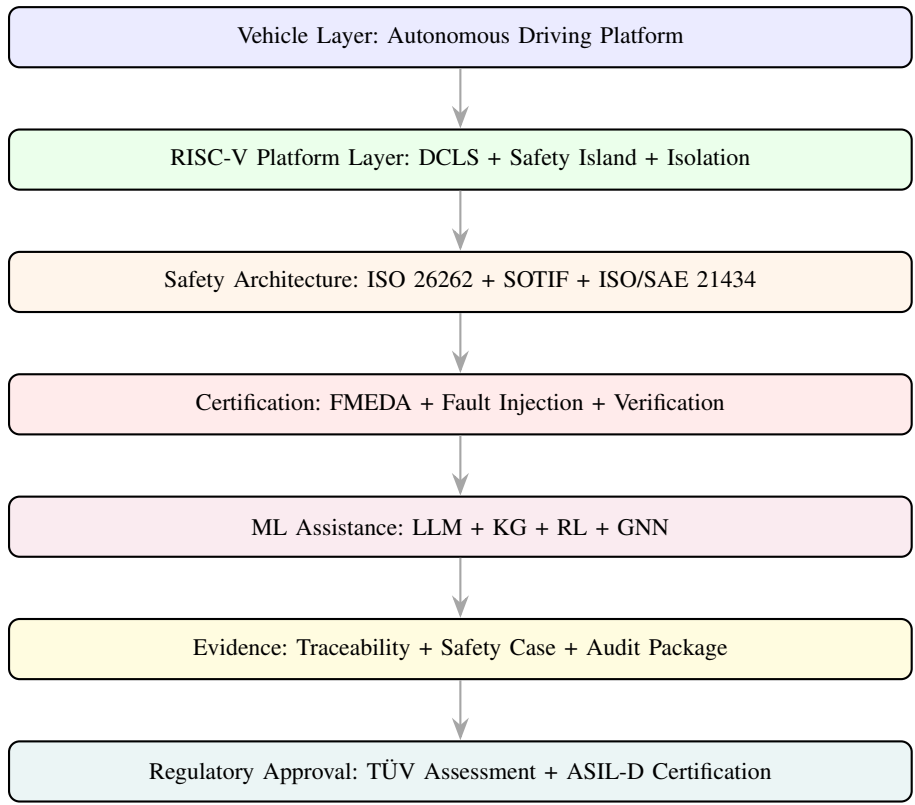
\begin{figure}[ht]
\centering
\begin{tikzpicture}[
  node distance=0.8cm,
  layer/.style={
    draw, rounded corners=4pt, minimum width=12cm,
    minimum height=0.8cm, align=center, font=\small,
    inner sep=4pt, thick
  },
  arrow/.style={-{Stealth[length=3mm]}, thick, gray!70}
]
\node[layer, fill=blue!8] (L1)
  {Vehicle Layer: Autonomous Driving Platform};
\node[layer, fill=green!8, below=of L1] (L2)
  {RISC-V Platform Layer: DCLS + Safety Island + Isolation};
\node[layer, fill=orange!8, below=of L2] (L3)
  {Safety Architecture: ISO~26262 + SOTIF + ISO/SAE~21434};
\node[layer, fill=red!8, below=of L3] (L4)
  {Certification: FMEDA + Fault Injection + Verification};
\node[layer, fill=purple!8, below=of L4] (L5)
  {ML Assistance: LLM + KG + RL + GNN};
\node[layer, fill=yellow!15, below=of L5] (L6)
  {Evidence: Traceability + Safety Case + Audit Package};
\node[layer, fill=teal!8, below=of L6] (L7)
  {Regulatory Approval: T\"{U}V Assessment + ASIL-D Certification};

\draw[arrow] (L1) -- (L2);
\draw[arrow] (L2) -- (L3);
\draw[arrow] (L3) -- (L4);
\draw[arrow] (L4) -- (L5);
\draw[arrow] (L5) -- (L6);
\draw[arrow] (L6) -- (L7);
\end{tikzpicture}
\caption{Seven-layer certification workflow for automotive
RISC-V autonomous driving platforms, integrating ISO~26262,
ISO~21448 (SOTIF), and ISO/SAE~21434.}
\label{fig:workflow}
\end{figure}

Figure~\ref{fig:workflow} presents the proposed seven-layer
certification workflow for automotive RISC-V platforms.

\section{Discussion and Limitations}
\label{sec:discussion}

\subsection{Research Implications}

Traditional academic architecture research focuses on IPC,
cache optimization, branch prediction, and microarchitectural
novelty. For automotive RISC-V, these considerations are
secondary. The highest-value research contribution lies in
\emph{certification infrastructure} rather than processor
performance.

The commercially successful RISC-V automotive platform may
not be the fastest. It may be the platform that is
\emph{easiest to certify}---the one with the strongest
compiler qualification package, the most comprehensive
T\"{U}V evidence package, the clearest safety manual, and
the fastest path to ASIL-D production deployment. This
represents a fundamental shift in how automotive processor
research should be funded, evaluated, and commercialized.

\subsection{Positioning as an Analytical Perspective}

This work does not propose a single isolated algorithmic
breakthrough. Instead, it provides a structured analytical
framework and expert-driven roadmap for enabling economically
superior certification of automotive-grade RISC-V platforms.
The certification economics model, RSMM, and ML-assisted
certification framework are intended as strategic evaluation
tools that structure the research space and identify
high-value problems. This positioning is appropriate for an
interdisciplinary domain characterized by emerging standards,
industrial transformation, and early-stage ecosystem
development.

\subsection{Limitations}

\begin{itemize}[leftmargin=*]
\item The analytical framework has not been empirically
validated through industrial case studies. The cost model
parameters require calibration against real certification
projects.

\item The RSMM levels are defined based on expert analysis
of current certification practice. As the RISC-V automotive
ecosystem matures, the model may require refinement.

\item ML-assisted certification faces significant barriers
to industrial adoption, including assessment body acceptance,
hallucination risk, and the meta-certification challenge of
validating ML-generated evidence
(Section~\ref{sec:ml_limits}).

\item The illustrative case study
(Section~\ref{sec:case_study}) demonstrates framework
applicability but does not constitute experimental validation.
\end{itemize}

\subsection{Future Work}

Priority future work includes: (i)~industrial pilot
deployment of the ML-assisted FMEDA generation pipeline on
a specific RISC-V implementation, (ii)~empirical calibration
of the certification cost model using data from real ASIL-D
projects, (iii)~development of formal acceptance criteria
for ML-generated certification evidence in collaboration
with assessment bodies, and (iv)~extension of the RSMM to
incorporate SOTIF and cybersecurity maturity dimensions.

\section{Conclusion}
\label{sec:conclusion}

RISC-V automotive adoption has entered a phase where ASIL-D
certification demonstrates technical feasibility
\cite{andes2025}. The strategic challenge is no longer
building safer processors but making safety certification
economically superior---especially for autonomous driving
systems that must simultaneously address ISO~26262,
ISO~21448 (SOTIF), and ISO/SAE~21434.

This paper contributes a structured analytical framework
comprising certification economics models, a five-level
RISC-V Safety Maturity Model, and an ML-assisted
certification methodology that maps LLMs with knowledge
graphs, reinforcement learning, and graph neural networks
to specific certification bottlenecks. The structured
Arm vs.\ RISC-V comparison demonstrates that RISC-V
provides structural advantages in ISA transparency, formal
verifiability, toolchain qualification portability, and
certification evidence reuse.

The future of automotive functional safety is not safer
CPUs---it is \emph{certifiable platforms}. The future of
certification is not manual compliance work---it is
\emph{structured, traceable, and intelligently assisted
certification infrastructure}. RISC-V is uniquely positioned
to become that platform.

The strongest outcome of this research direction is not an
academic paper but a deployable platform: \textbf{a certified,
secure, traceable, ASIL-D-ready RISC-V autonomous driving
platform}. This is where both academic impact and industrial
value converge.

\bibliographystyle{IEEEtran}
\bibliography{references}

\end{document}